# 数控技术中误差可控的三次样条 $C^2$ 插值

蔡泽威, 郑巍, 包益如, 楚海龙, 吴梦[*]

(南京理工大学数学与统计学院　南京　2100094)
(meng.wu@njust.edu.cn)

**摘　要**：在传统数控技术中构造符合 STEP 标准的 $C^2$ 连续的曲线时, 大多采用提高插值多项式次数的方式, 但这种方式会使得插值曲线出现龙格现象; 或是采取在端点处添加边值条件的办法, 而这样做往往会使的插值曲线的误差范围难以控制. 本文给出了一种 $C^2$ 连续的三次 B 样条曲线插值的方法, 该方法使得在插值多项式为三次的同时实现了插值曲线的 $C^2$ 连续性, 同时本文还研究了对应的误差控制方法.

**关键词**：三次样条; 加工路径; B 样条

**中图法分类号**: TP391.41　　　**DOI**: 10.3724/SP.J.1089.202*.

## Error Controlled Cubic Spline Interpolation in CNC Technology

Cai Zewei, Zheng Wei, Bao Yiru, Chu hailong, Wu Meng[*]

(School of Mathematics and Statistics, Nanjing University of Science and Technology, Nanjing　210094)

**Abstract:** Traditional CNC technology mostly uses the method of increasing the degree of interpolation polynomial when constructing $C^2$ continuous NURBS curves, but this often leads to the appearance of Runge phenomenon in interpolation curves; Alternatively, the method of adding boundary conditions at the endpoints can often make it difficult to control the error range of the interpolation curve. This article presents a $C^2$ continuous cubic B-spline curve interpolation method, which achieves $C^2$ continuity of the interpolation curve when the interpolation polynomial is cubic. At the same time, this article also studies the corresponding error control methods.

在数控加工中, 刀头轨迹经常表现为 $G^0$ 直线段的组合[1]. 使用 G1 或者 C1 曲线进行加工路径规划, 刀头的速度、加速度和加速度变化率不连续, 会造成机床剧烈振颤, 减少刀具和机床的使用寿命. 而加工路径的 $C^2$ 连续性可以避免这些问题.

符合 STEP 标准的曲线曲面重新插值或重新拟合技术是计算机辅助几何设计(CAGD)技术的重要组成部分[3]. 尤其是在加工路径拟合中, 往往需要将 offset 曲线转化为符合 STEP 标准的曲线. 因此符合 STEP 标准的 $C^2$ 插值曲线生成是数控技术中重要的课题.

国内外已经有许多研究人员对这一问题开展了深入的研究, 并提出了各种不同的解决办法. 例如: 张勇提出了一种通过五次 B 样条曲线生成四

阶几何连续的刀具路径平滑算法，并给出了相应的突变平滑进给速度的调度方案[1]；乔志峰提出了一种新的动态偏移误差控制方法，该方法是对 Bézier 曲线快速细分方法的扩展，能够在用户要求的偏移精度下实现 Bézier 或 NURBS 曲线的快速偏移[3]；Soo Won Kim 给出了一种基于二次 Bézier 曲线圆形逼近的曲率连续有理 offset 曲线逼近方法[4]；Raine 提出了利用矩阵加权 NURBS 曲线对点序列与点上的法线或切线进行拟合和插值的方法[6]；李军成构造了一种带参数的自动满足 $C^2$ 连续的五次 Hermite 插值样条以及另一种带参数的 $C^3$ 连续分段七次 Hermite 插值样条[11]；韩江给出了一种基于轮廓关键点的 B 样条插值[12]；冯仁忠提出了在局部直线段、尖点区域构造 $C^2$ 连续的样条曲线的方法[13]；陈良骥针对计算机数控编程阶段生成的海量离散刀头位置数据，在满足预设精度的条件下，给出了一种采用曲率自适应选取特征点非均匀有理 B 样条曲线插值[14]；李军成提出了一种带参数的五次 Hermite 插值样条的方法，在固定插值条件时，可用所带的形状参数对曲线进行调控[17]。

上述文献在构造符合 STEP 标准的 $C^2$ 曲线时大致可归结为两类：一是提高插值多项式次数的，但这种方式会使得插值曲线出现龙格现象；二是在端点处添加一定量的边值条件，而这种方法会使的插值曲线的误差难以控制。

本文给出了一种误差可控的 $C^2$ 连续三次 B 样条曲线插值的方法，该方法使得在插值多项式为三次的同时实现了插值曲线的 $C^2$ 连续性，同时本文还研究了对给定误差插值曲线自适应步长的选取方法。具体而言，本文的贡献如下：

（1）通过增加节点的方式，引入充分多的自由度，构造满足 STEP 标准的低次(三次) $C^2$ 连续的 B 样条插值曲线；

（2）从理论上分析了该插值方法的存在唯一性以及插值逼近的逼近误差；

（3）在给定误差情况下，我们讨论了自适应步长的选取。特别地，对于直线路径，该插值方法对步长无限制，即可选择较大步长以减少计算量。

（4）根据理论分析，我们给出了实际加工路径的数值算例。

本文的误差可控的三次样条 $C^2$ 插值方法按照以下步骤对原路径曲线进行插值的。首先根据给定误差以及原路径信息自适应选择插值步长。把原曲线分解为定义在 $[t_i, t_{i+1}]$ 上的小曲线段；其次对于每个小曲线段进行 $C^2$ 的三次 B 样条插值。本文提出的方法可以确保该插值整体 $C^2$ 连续性。为了便于呈现结果，后文结构如下。在第一部分中，我们考虑对函数在区间端点上特殊的 $C^2$ 三次 B 样条插值；在第二部分中，我们对这种插值进行了误差分析，同时基于误差分析，我们讨论了给定的误差对原曲线进行自适应分解，给出分解的参数步长选择公式；在第三部分中，我们给出在加工路径生成中的数值算例；最后在第四部分中对本文的工作进行总结。

## 1　端点插值的 $C^2$ 三次 B 样条

首先我们考虑函数在单位区间上的一种特殊的 $C^2$ 三次 B 样条端点插值；其次通过参数变换，把一般区间 $[t_i, t_{i+1}]$ 上的函数插值问题转化为单位区间上的插值，得到 $[t_i, t_{i+1}]$ 区间上的 $C^2$ 三次 B 样条端点插值。

### 1.1　单位区间上 $C^2$ 三次 B 样条端点插值

**定理 1**：设 $y = f(t), t \in [0,1]$，考虑节点向量

$$U = \left\{0,0,0,0,\frac{1}{3},\frac{2}{3},1,1,1,1\right\},$$

设 $b(t)$ 是定义于 $U$ 上的三次 $B$ 样条，且满足

$$\frac{\partial^k b}{\partial t^k}(0) = \frac{\partial^k f}{\partial t^k}(0),$$

$$\frac{\partial^k b}{\partial t^k}(1) = \frac{\partial^k f}{\partial t^k}(1),$$

其中 $k = 0,1,2$，则 $b(t)$ 存在且唯一.

即：$$b(t) = \sum_{i=0}^{5} \lambda_i N_{i,3}(t),$$

其中：

$$\begin{cases} \lambda_0 = f_{0,0}, \\ \lambda_1 = \frac{1}{9} f_{0,1} + f_{0,0}, \\ \lambda_2 = \frac{1}{27} f_{0,2} + \frac{1}{3} f_{0,1} + f_{0,0}. \end{cases} \quad (1.1)$$

$$\begin{cases} \lambda_3 = \frac{1}{27} f_{1,2} - \frac{1}{3} f_{1,1} + f_{1,0}, \\ \lambda_4 = f_{1,0} - \frac{1}{9} f_{1,1}, \\ \lambda_5 = f_{1,0}. \end{cases} \quad (1.2)$$

定理 1 的证明见附录第一部分. □

为了在第 2 节中进行误差分析，这里我们计算 $b''(t)$. 由于 $b''(t)$ 是 $C^0$ 且为一次分片多项式，故 $b''(t)$ 完全由 $t = 0, \frac{1}{3}, \frac{2}{3}, 1$. 处的函数值决定. 其中 $b''(0) = f''(0)$，$b''(1) = f''(1)$，下面我们计算 $b''(\frac{1}{3})$，$b''(\frac{2}{3})$.

$$b''\left(\frac{1}{3}\right) = \delta_1 N_{1,1}\left(\frac{1}{3}\right) = \delta_1 = \\ -9 f_{0,0} - 6 f_{0,1} - \frac{5}{6} f_{0,2} + 9 f_{1,0} - 3 f_{1,1} + \frac{1}{3} f_{1,2}, \quad (1.3)$$

$$b''\left(\frac{2}{3}\right) = \delta_2 N_{2,1}\left(\frac{2}{3}\right) = \delta_2 = \\ 9 f_{0,0} + 3 f_{0,1} + \frac{1}{3} f_{0,2} - 9 f_{1,0} + 6 f_{1,1} - \frac{5}{6} f_{1,2}. \quad (1.4)$$

由 $b(t)$ 的存在唯一性，我们可以得到如下推论.

**推论 1**：当 $y = f(t), t \in [0,1]$ 为三次多项式，则
$$b(t) = f(t),$$
其中 $b(t)$ 在定理 1 中定义.

### 1.2 一般的区间 $[t_i, t_{i+1}]$ 上 $C^2$ 三次 $B$ 样条端点插值

设 $F(t)$ 为定义于 $[t_i, t_{i+1}]$ 的函数，$B(t)$ 为定义于
$$U = \{t_i, t_i, t_i, t_i, t_i + h_i, t_i + 2h_i, t_{i+1}, t_{i+1}, t_{i+1}, t_{i+1}\} \text{ 的}$$
$C^2$ 三次 $B$ 样条，且
$$B^{(k)}(t_l) = F^{(k)}(t_l) \quad (1.5)$$
其中 $l = i, i+1$，$k = 0, 1, 2$，$h_i = \frac{t_{i+1} - t_i}{3}$. 下面我们根据 1.1 节结果计算 $B(t)$ 的具体公式.

令 $u = \frac{t - t_i}{t_{i+1} - t_i}$,

考虑 $b(u) = B((t_{i+1} - t_i)u + t_i)$，则其满足:

$$\begin{cases} b(0) = B(t_i) = F(t_i) := f_{0,0}, \\ b'(0) = \frac{\partial B}{\partial t}(t_i) \frac{\partial t}{\partial u}\bigg|_{u=0} = (t_{i+1} - t_i) F'(t_i) := f_{0,1}, \\ b''(0) = \frac{\partial}{\partial u}\left(\frac{\partial t}{\partial u} \frac{\partial B}{\partial t}\right)\bigg|_{u=0} = (t_{i+1} - t_i)^2 f''(t_i) := f_{0,2}, \\ b(1) = B(t_{i+1}) = F(t_{i+1}) := f_{1,0}, \\ b'(1) = B'(t_{i+1})(t_{i+1} - t_i) = (t_{i+1} - t_i) F'(t_{i+1}) := f_{1,1}, \\ b''(1) = B''(t_{i+1})(t_{i+1} - t_i)^2 = (t_{i+1} - t_i)^2 F''(t_{i+1}) := f_{1,2}. \end{cases} \quad (1.6)$$

由定理 1，存在唯一 $b(u)$ 满足 (1.6). 我们考虑 $B(t) = b\left(\frac{t - t_i}{t_{i+1} - t_i}\right)$ 在 $t = t_i, t_{i+1}$ 处的 0, 1, 2 阶导数，根据 (1.5) 和 (1.6)，其为所求 B 样条.

对于定义于 $[a, b]$ 上的 $F(t)$，对 $[a, b]$ 进行分割，
$$T: a = t_0 < t_1 < \cdots < t_n = b,$$
在 $[t_i, t_{i+1}]$ 上我们均可构造 $C^2$ 三次 $B$ 样条插值于端点 $t_i, t_{i+1}$，即满足 (1.5)，记为 $B_i(t)$. 因此，在 $[a, b]$ 上可构造 $C^2$ 三次 $B$ 样条 $B(t)$ 使得
$$B(t)|_{[t_i, t_{i+1}]} = B_i(t),$$
即,
$$B^{(k)}(t_l) = F^{(k)}(t_l), k = 0, 1, 2, l = 0, 1, \cdots n.$$
即，$B(t)$ 是定义在 $[a, b]$ 上在 $t_i$ 处插值于 $F(t)$ 的 $C^2$ 三次 $B$ 样条. 下一节中我们将对这种插值进行误差分析并在给定误差的情况下自适应参数步长的选取条件.

### 2、误差分析和自适应参数步长的选取

**定理 2**：设 $F(t), t \in [t_i, t_{i+1}]$，$B(t)$ 是定义于
$$U = \{t_i, t_i, t_i, t_i, t_i + h_i, t_i + 2h_i, t_{i+1}, t_{i+1}, t_{i+1}, t_{i+1}\},$$
的 $C^2$ 三次 B 样条，$h_i = \frac{1}{3}(t_{i+1} - t_i)$，且 $B(t)$ 满足
$$B^{(k)}(t_l) = F^{(k)}(t_l), \quad k = 0, 1, 2; l = i, i+1$$
考虑误差函数
$$R(t) = F(t) - B(t), \quad t \in [t_i, t_{i+1}]$$
存在 $\xi_{i(t)} \in [t_i, t_{i+1}]$ 使得
$$R(t) = -\frac{R''(\xi_{i(t)})}{2}(t - t_i)(t_{i+1} - t),$$
定理 2 的证明见附录第二部分.

**定理 3：** 条件如定理 2 中所述，则对于 $[t_i, t_{i+1}]$

$$|R(t)| \leq \frac{M_i}{8} h_i^2, \quad (2.1)$$

其中，

$$M_i = 7 \max_{t \in [t_i, t_{i+1}]} |F''(t)| +$$

$$\max \left\{ \frac{5}{6}|F''(t_i)| + \frac{1}{3}|F''(t_{i+1})|, \frac{1}{3}|F''(t_i)| + \frac{5}{6}|F''(t_{i+1})| \right\}, (2.2)$$

是只依赖 $F(t)$ 和 $[t_i, t_{i+1}]$ 的常数.

定理 3 的证明见附录第三部分

对于给定误差 $d$ 以及定义于 $[a,b]$ 函数 $F(t)$，考虑对 $[a,b]$ 的剖分

$$T: a = t_0 < t_1 < t_2 < \cdots < t_n = b$$

使得

$$|F(t) - B(t)| \leq d \ (\forall \ t \in [a,b]),$$

则

$$h_i \leq \sqrt{\frac{8d}{M_i}}. \quad \square$$

下面我们考虑在 $[a,b]$ 上步长的选择，记 $R(t) = F(t) - B(t)$，则：

$$|R(t)| \leq \max_{i=0,1,\cdots n-1} \max_{t \in [t_i, t_{i+1}]} |R(t)| \leq \frac{M}{8} \max_{i=0,1,\cdots n-1} (t_{i+1} - t_i)^2$$

其中，$M = \max_{i=0,1,\cdots n-1} M_i$，$M_i$ 如(2.2)定义

记：

$$h = \max_{i=0,1,\cdots n-1} (t_{i+1} - t_i),$$

则若：

$$\frac{M}{8} h^2 \leq d \quad (2.3)$$

则 $|R(t)| \leq d$，由（2.3）得

$$h \leq \sqrt{\frac{8d}{M}}.$$

对于平面上参数曲线 $(x(t), y(t))$，分别对 $x$ 方向和 $y$ 方向的函数进行 $C^2$ 插值，记此三次 $C^2$ $B$ 样条插值为 $(B_1(t), B_2(t))$，对于给定误差 $d$，使得

$$(x(t) - B_1(t))^2 + (y(t) - B_2(t))^2 \leq d^2 \quad (2.4)$$

对于 $x$ 方向，根据(2.3)，存在常值 $Mx$，使得

$$|x(t) - B_1(t)| \leq \frac{Mx}{8} h^2,$$

若 $\left(\frac{Mx}{8} h^2\right)^2 + \left(\frac{My}{8} h^2\right)^2 \leq d^2$，则(2.4)成立，此时，

$$h \leq \frac{\sqrt{8d}}{\left(M_x^2 + M_y^2\right)^{\frac{1}{4}}}.$$

同理，对于空间参数曲线 $(x(t), y(t), z(t))$，

$$h \leq \frac{\sqrt{8d}}{\left(M_x^2 + M_y^2 + M_z^2\right)^{\frac{1}{4}}}.$$

## 3. 数值算例

本节中我们选择三个例子:

**例 1：** 在这个例子中，我们选取三次多项式

$$y = x^3 + 31x^2 - 6x + 119$$

作为插值对象. 根据推论 1 知，该插值为精确插，即插值曲线(如图 1)与原多项式完全重合.

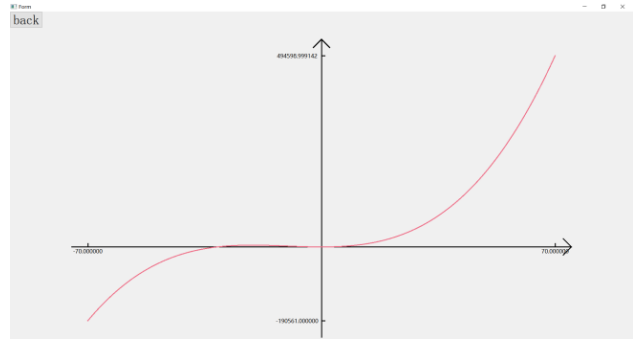

图 1 例 1 中三次多项式的插值函数

**例 2：** 在这个例子中，我们选择机械加工中由 Bézier 曲线段拼接形成的复杂加工路径(图 2 中红色曲线). 对其 offset 曲线使用本文中的可控误差光滑插值方法进行插值可得刀具加工轨迹曲线(图 2 中黄色曲线).

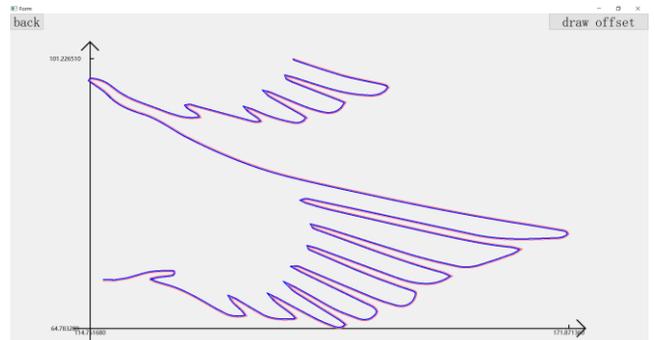

图 2 例 2 中复杂加工路径(红色)以及其 offset 曲线的可控误差插值曲线(黄色)，其中算法中尖角处采用方角补偿.

**例 3**：对于机械加工中由 Bézier 曲线段拼接形成的复杂曲线的 offset 曲线具有不同的曲线弯曲情况(如图 3 所示). 在控制误差约束下，我们可以采用自适应选取其较大的插值区间长度从而减少计算量.

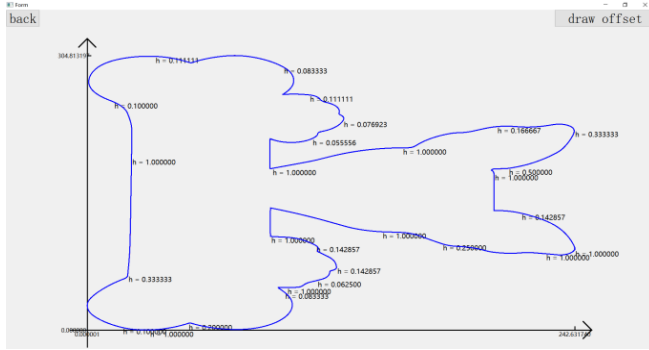

图 3 例 3 中复杂加工路径的 offset 曲线.当误差给定时，自适应步长 h 的选择。

特别地，对于直线段(2.2)式中的 $M_i$ 均为 0，因此对直线段的插值无最大步长限制. 在算法中我们选取参数最大区间长度，即

$$h = 1.$$

## 4 结 语

本文中我们构造了 $C^2$ 连续的三次 B 样条插值曲线，同时对这种插值方法进行了误差分析. 并基于误差分析研究了给定误差情况下自适应步长的选取，最后给出了在多项式函数插值、复杂 offset 曲线插值以及自适应步长选取算例. 未来我们将进一步使用该理论用于其他工程场景.

附录：

**定理 1**: 设 $y=f(t), t\in[0,1]$, 考虑节点向量

$$U=\left\{0,0,0,0,\frac{1}{3},\frac{2}{3},1,1,1,1\right\},$$

设 $b(t)$ 是定义于 $U$ 上的三次 $B$ 样条, 且满足

$$\frac{\partial^k b}{\partial t^k}(0)=\frac{\partial^k f}{\partial t^k}(0),$$

$$\frac{\partial^k b}{\partial t^k}(1)=\frac{\partial^k f}{\partial t^k}(1),$$

其中 $k=0,1,2$, 则 $b(t)$ 存在且唯一.

证明: 记 $N_{i,3}(t)$ 为定义于 $U$ 的基函数, $i=0,1,2,\cdots,5$.

令

$$b(t)=\sum_{i=0}^{5}\lambda_i N_{i,3}(t),$$

记

$$U_1=\left\{0,0,0,\frac{1}{3},\frac{2}{3},1,1,1\right\}=\{u_0,u_1,\cdots,u_3,u_4,u_5,\cdots,u_7\},$$

$$b'(t)=3\sum_{i=0}^{4}N_{i,2}(t)\cdot\frac{\lambda_{i+1}-\lambda_i}{u_{i+3}-u_i},$$

$$b'(t)=3\left\{\frac{\lambda_1-\lambda_0}{\frac{1}{3}}N_{0,2}(t)+\frac{\lambda_2-\lambda_1}{\frac{2}{3}}N_{1,2}(t)+\right.$$
$$\left.(\lambda_3-\lambda_2)N_{2,2}(t)+\frac{\lambda_4-\lambda_3}{\frac{2}{3}}N_{3,2}(t)+\frac{\lambda_5-\lambda_4}{\frac{1}{3}}N_{4,2}(t)\right\},$$

其中 $N_{i,2}(t)$ 定义于 $U$ 的二次 $B$ 样条基函数. 即,

$$b'(t)=9(\lambda_1-\lambda_0)N_{0,2}(t)+\frac{9}{2}(\lambda_2-\lambda_1)N_{1,2}(t)$$
$$+3(\lambda_3-\lambda_2)N_{2,2}(t)+\frac{9}{2}(\lambda_4-\lambda_3)N_{3,2}(t)$$
$$+9(\lambda_5-\lambda_4)N_{4,2}(t)$$
$$\triangleq\sum_{i=0}^{4}v_i N_{i,2}(t)$$

记 $U_2=\left\{0,0,\frac{1}{3},\frac{2}{3},1,1\right\}\triangleq\{u_0,u_1,u_2,u_3,u_4,u_5\}$,

则:

$$b''(t)=2\sum_{i=0}^{3}N_{i,1}(t)\frac{v_{i+1}-v_i}{u_{i+2}-u_i}$$
$$=6(v_1-v_0)N_{0,1}(t)+3(v_2-v_1)N_{1,1}(t)$$
$$+3(v_3-v_2)N_{2,1}(t)+6(v_4-v_3)N_{3,1}(t)$$
$$=(54\lambda_0-81\lambda_1+27\lambda_2)N_{0,1}(t)$$
$$+\left(\frac{27}{2}\lambda_1-\frac{45}{2}\lambda_2+9\lambda_3\right)N_{1,1}(t)$$
$$+\left(9\lambda_2-\frac{45}{2}\lambda_3+\frac{27}{2}\lambda_4\right)N_{2,1}(t)$$
$$+(27\lambda_3-81\lambda_4+54\lambda_5)N_{3,1}(t)$$
$$\triangleq\sum_{i=0}^{3}\delta_i N_{i,1}(t)$$

由 $\dfrac{\partial^k b(0)}{\partial t^k}=\dfrac{\partial^k f(0)}{\partial t^k}$ $(k=0,1,2)$, 得

$$\begin{cases}\lambda_0=f(0),\\ 9(\lambda_1-\lambda_0)=f'(0),\\ 54\lambda_0-81\lambda_1+27\lambda_2=f''(0).\end{cases}$$

记 $f_{0,k}=\dfrac{\partial^k f(0)}{\partial t^k}$, 得

$$\begin{cases}\lambda_0=f_{0,0},\\ \lambda_1=\dfrac{1}{9}f_{0,1}+f_{0,0},\\ \lambda_2=\dfrac{1}{27}f_{0,2}+\dfrac{1}{3}f_{0,1}+f_{0,0}.\end{cases} \quad (1.1)$$

同理, 由 $\dfrac{\partial^k b(1)}{\partial t^k}=\dfrac{\partial^k f(1)}{\partial t^k}(k=0,1,2)$ 得

$$\begin{cases}\lambda_5=f_{1,0},\\ 9(\lambda_5-\lambda_4)=f_{1,1},\\ 27\lambda_3-81\lambda_4+54\lambda_5=f_{1,2}.\end{cases}$$

其中,

$$f_{1,k}=\frac{\partial^k f(1)}{\partial t^k},$$

即,

$$\begin{cases}\lambda_3=\dfrac{1}{27}f_{1,2}-\dfrac{1}{3}f_{1,1}+f_{1,0},\\ \lambda_4=f_{1,0}-\dfrac{1}{9}f_{1,1},\\ \lambda_5=f_{1,0}.\end{cases} \quad (1.2)$$

故 $b(t)$ 存在且系数由(1.1)(1.2)唯一确定. □

**定理 2**：设 $F(t), t \in [t_i, t_{i+1}]$，$B(t)$ 是定义于
$U = \{t_i, t_i, t_i, t_i + h_i, t_i + 2h_i, t_{i+1}, t_{i+1}, t_{i+1}, t_{i+1}\}$，
的 $C^2$ 三次 B 样条，$h_i = \frac{1}{3}(t_{i+1} - t_i)$，且 $B(t)$ 满足
$B^{(k)}(t_l) = F^{(k)}(t_l)$，$k = 0,1,2; l = i, i+1$
考虑误差函数
$R(t) = F(t) - B(t)$，$t \in [t_i, t_{i+1}]$
存在 $\xi_{i(t)} \in [t_i, t_{i+1}]$ 使得
$$R(t) = -\frac{R''(\xi_{i(t)})}{2}(t - t_i)(t_{i+1} - t),$$

证明：构造
$$F(x) = R(x) - R(t)\frac{(x - t_i)(t_{i+1} - x)}{(t - t_i)(t_{i+1} - t)} \in C^2(t_i, t_{i+1}),$$
且 $F(t_i) = 0$，$F(t_{i+1}) = 0$.

故 $F(x)$ 在 $[t_i, t_{i+1}]$ 上连续，$(t_i, t_{i+1})$ 内可导，则存在 $\eta_i \in [t_i, t_{i+1}]$，使得
$F(\eta_i) = 0.$

另外，
$$F'(x) = R'(x) - R(t)\frac{-2x + (t_{i+1} - t_i)}{(t - t_i)(t_{i+1} - t)} \in C^i(t_i, t_{i+1})$$
$F'(t_i) = 0$，$F'(t_i) \in C^0(t_i, t_{i+1})$.

故，存在 $\xi_{i(t)} \in [t_i, t_{i+1}]$，使得
$F''(\xi_i) = 0.$

而 $F''(x) = R''(x) + \frac{2R(t)}{(t - t_i)(t_{i+1} - t)}$

故 $F''(\xi_i) = R''(\xi_i) + \frac{2R(t)}{(t - t_i)(t_{i+1} - t)} = 0$，$t \in (t_i, t_{i+1})$

故，
$$R(t) = -\frac{R''(\xi_i)}{2}(t - t_i)(t_{i+1} - t). \quad (2.1)$$

由 $R(t)$，$R''(t)$ 的连续性可得(2.1)在 $t \in (t_i, t_{i+1})$ 上均成立.

**定理 3**：条件如定理 2 中所述，则对于 $[t_i, t_{i+1}]$
$$|R(t)| \leq \frac{M_i}{8} h_i^2, \quad (2.2)$$
其中，
$$M_i = 7 \max_{t \in [t_i, t_{i+1}]} |F''(t)|$$
$$+ \max\left\{\frac{5}{6}|F''(t_i)| + \frac{1}{3}|F''(t_{i+1})|, \frac{1}{3}|F''(t_i)| + \frac{5}{6}|F''(t_{i+1})|\right\},$$
是只依赖 $F(t)$ 和 $[t_i, t_{i+1}]$ 的常数.

证明：由(2.1)可得 $|R(t)|$ 的一个上界
$$|R(t)| \leq \frac{|R''(\xi)|}{8}(t_{i+1} - t_i)^2. \quad (2.3)$$

我们进一步分析 $R''(t) = F''(t) - B''(t)$，其中，$B''(t)$ 是 $C^0$ 的分片线性函数，故

$$|R''(t)| \leq \max\{|B''(t_i)|, |B''(t_{i+1})|, |B''(t_i + h_i)|, |B''(t_i + 2h_i)|\},$$
$B''(t_i) = F''(t_i)$，$B''(t_{i+1}) = F''(t_{i+1})$，
$$B''(t_i + h_i) = \frac{1}{(t_{i+1} - t_i)^2} b''\left(\frac{1}{3}\right),$$
$$B''(t_i + 2h_i) = \frac{1}{(t_{i+1} - t_i)^2} b''\left(\frac{2}{3}\right).$$

由(2.1)和(2.2)式，
$$b''\left(\frac{1}{3}\right) = -9f_{0,0} - 6f_{0,1} - \frac{5}{6}f_{0,2} + 9f_{1,0} - 3f_{1,1} + \frac{1}{3}f_{1,2}$$
$$b''\left(\frac{2}{3}\right) = 9f_{0,0} + 3f_{0,1} + \frac{1}{3}f_{0,2} - 9f_{1,0} + 6f_{1,1} - \frac{5}{6}f_{1,2}$$

其中，$f_{i,j}(i = 0,1, j = 0,1,2)$ 如（1.6）式中定义，因此，

$$|R''(t)| = |F''(t) - b''(t)| \leq$$
$$|F''(t)| + |b''(t)| \leq \max_{t \in [t_i, t_{i+1}]} |F''(t)| +$$
$$\max\left\{|F''(t_i)|, |F''(t_{i+1})|, \frac{|b''(\frac{1}{3})|}{(t_{i+1} - t_i)^2}, \frac{|b''(\frac{2}{3})|}{(t_{i+1} - t_i)^2}\right\} \quad (2.4)$$

记 $h_i = t_{i+1} - t_i$，

$$\left|b''\left(\frac{1}{3}\right)\right| = \frac{-9f_{0,0} - 6f_{0,1} - \frac{5}{6}f_{0,2} + 9f_{1,0} - 3f_{1,1} + \frac{1}{3}f_{1,2}}{h_i^2}$$

$$= \frac{-9F(t_i) - 6F'(t_i)h_i - \frac{5}{6}F''(t_i)h_i^2 + 9F(t_{i+1})}{h_i^2}$$

$$+ \frac{-3F'(t_{i+1})h_i + \frac{1}{3}F''(t_{i+1})h_i^2}{h_i^2}$$

$$= 9\frac{F(t_{i+1}) - F(t_i)}{h_i^2} - \frac{6F'(t_i) + 3F'(t_{i+1})}{h_i} - \frac{5}{6}F''(t_i) + \frac{1}{3}F''(t_{i+1})$$

$$= 6\frac{F'(\xi_i) - F'(t_i)}{h_i} - 3\frac{F'(t_{i+1}) - F'(\xi_i)}{h_i} - \frac{5}{6}F''(t_i) + \frac{1}{3}F''(t_{i+1})$$

$$= 6F''(\eta_1)\frac{\xi_i - t_i}{h_i} - 3F''(\eta_2)\frac{t_{i+1} - \xi_i}{h_i} - \frac{5}{6}F''(t_i) + \frac{1}{3}F''(t_{i+1}),$$

$$\frac{\left|b''\left(\frac{1}{3}\right)\right|}{h_i^2} = 6F''(\eta_1)\frac{\xi_i - t_i}{h_i} - 3F''(\eta_2)\frac{t_{i+1} - \xi_i}{h_i}$$

$$- \frac{5}{6}F''(t_i) + \frac{1}{3}F''(t_{i+1})$$

因此,

$$\frac{\left|b''\left(\frac{1}{3}\right)\right|}{h_i^2} \leq 6\max_{t \in [t_i, t_{i+1}]}|F''(t)| + \frac{5}{6}|F''(t_i)| + \frac{1}{3}|F''(t_{i+1})| \quad (2.5)$$

同理,

$$\frac{\left|b''\left(\frac{2}{3}\right)\right|}{h_i^2} = 6F''(\eta_1)\frac{t_{i+1} - \xi_i}{h_i} - 3F''(\eta_2)\frac{\xi_i - t_i}{h_i} -$$

$$\frac{5}{6}F''(t_{i+1}) + \frac{1}{3}F''(t_i) \quad (2.6)$$

其中, $\xi_i \in (t_i, t_{i+1})$, $\eta_1 \in (t_i, \xi_i)$, $\eta_2 \in (\xi_i, t_{i+1})$, 因此,

$$\frac{\left|b''\left(\frac{2}{3}\right)\right|}{(t_{i+1} - t_i)^2} \leq 6\max_{t \in [t_i, t_{i+1}]}|F''(t)| + \frac{5}{6}|F''(t_{i+1})| + \frac{1}{3}|F''(t_i)| \quad (2.7)$$

由(2.4)、(2.5)、(2.7)得, 当 $t \in [t_i, t_{i+1}]$ 时,

$$|R''(t)| \leq 7\max_{t \in [t_i, t_{i+1}]}|F''(t)|$$

$$+ \max\left\{\frac{5}{6}|F''(t_i)| + \frac{1}{3}|F''(t_{i+1})|, \frac{1}{3}|F''(t_i)| + \frac{5}{6}|F''(t_{i+1})|\right\} \quad (2.8)$$

结合(2.3)可得,

$$|R(t)| \leq \frac{M_i}{8}h_i^2,$$

其中,

$$M_i = 7\max_{t \in [t_i, t_{i+1}]}|F''(t)| +$$

$$\max\left\{\frac{5}{6}|F''(t_i)| + \frac{1}{3}|F''(t_{i+1})|, \frac{1}{3}|F''(t_i)| + \frac{5}{6}|F''(t_{i+1})|\right\}, (2.9)$$

是只依赖 $F(t)$ 和 $[t_i, t_{i+1}]$ 的常数.